\documentclass[12pt,a4paper]{cibb}

\makeatletter
\providecommand{\@ordinalM}[2]{#1}
\makeatother

\usepackage{subcaption,graphicx}
\usepackage{amsmath,amsfonts,latexsym,amssymb,euscript,xr}
\usepackage{booktabs}
\usepackage[nodayofweek]{datetime}
\usepackage{hyperref}
\usepackage{fmtcount}
\usepackage[english]{datenumber}
\usepackage[absolute]{textpos}
\usepackage{subcaption}
\usepackage{graphicx}

\usepackage[table]{xcolor}
\usepackage{color,colortbl,tabularx}

\usepackage[english]{babel}
\usepackage[protrusion=true,expansion=true]{microtype}
\usepackage{amsmath,amsfonts,amsthm}
\usepackage{pifont}
\usepackage{tikz}
\usetikzlibrary{arrows.meta, positioning}

\definecolor{LightBlue}{rgb}{0.88,0.9,0.9}

\title{\Large $\ $\\ \bf Katz Kernel Based Stratification of Breast Cancer Patients}

\author{Francesco Cellitti$^1$, Lorenzo Di Rocco$^{1}$ and Umberto Ferraro Petrillo$^{1}$}
\address{\footnotesize $\ $\\$^1$ Sapienza University of Rome,
Rome, Italy. \\

\bigskip
ORCID codes: FC 0009-0000-5634-7431; LDR 0000-0001-8744-7048; UFP 0000-0002-4308-5126.
\bigskip
\newline
$^*$corresponding author: francesco.cellitti@uniroma1.it
}

\abstract{\small signal mechanistic networks, graph kernel, graph clustering. \normalsize
\\[17pt]
{\bf Abstract (max 250 words).} The integration of mechanistic signaling networks into clinical research has become a cornerstone of precision oncology. By capturing the complex biochemical interactions within a cell, these networks provide a functional map of disease progression that goes beyond simple genomic mutations. Consequently, the development of specific signaling models of patients has significant relevance, allowing for a more granular representation of individual disease profiles. In this setting, there is a growing need for robust methodologies capable of comparing graph based pathway structures to identify similarities and differences across patient cohorts. Such comparisons are essential for the identification of clinically meaningful subgroups, and the discovery of novel biomarkers. Standard graph kernels, while mathematically rigorous, often remain too generic for this purpose. They frequently fail to capture features specific to gene signaling networks, resulting in suboptimal clustering performance. In this paper, we present a domain specific graph kernel designed for mechanistic signaling networks. Unlike generic approaches, our method explicitly incorporates biological constraints and gene interaction patterns into the kernel formulation, enabling a more accurate characterization of both topological structure and functional behavior. We assess its performance on breast cancer datasets, focusing on its ability to support disease subtyping. Our results suggest that our approach can effectively contribute to patient stratification.

}

\begin{document}

\renewcommand{\thefootnote}{}
\footnotetext{\small{Article version: \datedate $\;$ h\currenttime  $\;$ CET}}

\thispagestyle{myheadings}
\pagestyle{myheadings}
\markright{\tt Proceedings of CIBB 2026}

\section{Introduction}
\label{sec:SCIENTIFIC-BACKGROUND}

While traditional transcriptomic approaches provide snapshots of gene expression, they often fail to capture the functional dynamics of cellular signaling. A simple list of differentially expressed genes ignores the topological context in which these genes operate and the flow of information across biochemical pathways.

Whilst Hofree’s Network Based Stratification (NBS, \cite{hofree2013network}) model uses generic, static protein to protein interaction networks, our approach employs mechanistic signalling networks. This allows us to capture specific biochemical dynamics, rather than merely the physical proximity between proteins.

In this work, we propose a framework for patient stratification based on Graph Kernels applied to signaling networks. We represent each patient as a directed graph where nodes represent both the identity and the functional state of a gene in an expanded $2N \times 2N$ space. By employing the Katz centrality measure, we capture the global influence of each gene within the network, accounting for both direct and long range interactions.

The similarity between patients is computed using the Frobenius inner product (Trace) of their respective Katz diffusion matrices. This approach defines a formal Mercer Kernel, allowing us to project patients into a high dimensional Hilbert space where similarity reflects mechanistic alignment rather than mere statistical overlap. We demonstrate the effectiveness of this method on a breast cancer dataset, showing that the formation of linkage between patient specific signaling topologies through the Katz Kernel leads to a robust stratification.

\section{Methods}
\label{sec:methods}

Firstly let's define mechanistic signaling networks as a directed graph $\mathcal{G}$ made of different genes, described as vertices $\mathcal{V}$. Each interaction between genes is characterized using a directed edge $\mathcal{E}$ where the sense of direction means that the source gene activates or inhibits the target gene (respectively considered as -1 and 1). We define this property as the node state $s_v = \{-1, +1\}$. 

The idea is to compare the graph using an adjacency matrix $A$ made of all the possible genes. Since the activation property of the gene is fundamental (a gene B that activates C is completely different from a gene B that inhibits C), the adjacency matrix is built on a dimension $2N \times 2N$. Each gene is described in the matrix with two rows, one for each activation as shown in the left plot of Figure \ref{example_katz}.

\begin{figure}[h]
 \begin{center}
 \includegraphics[width=\textwidth]{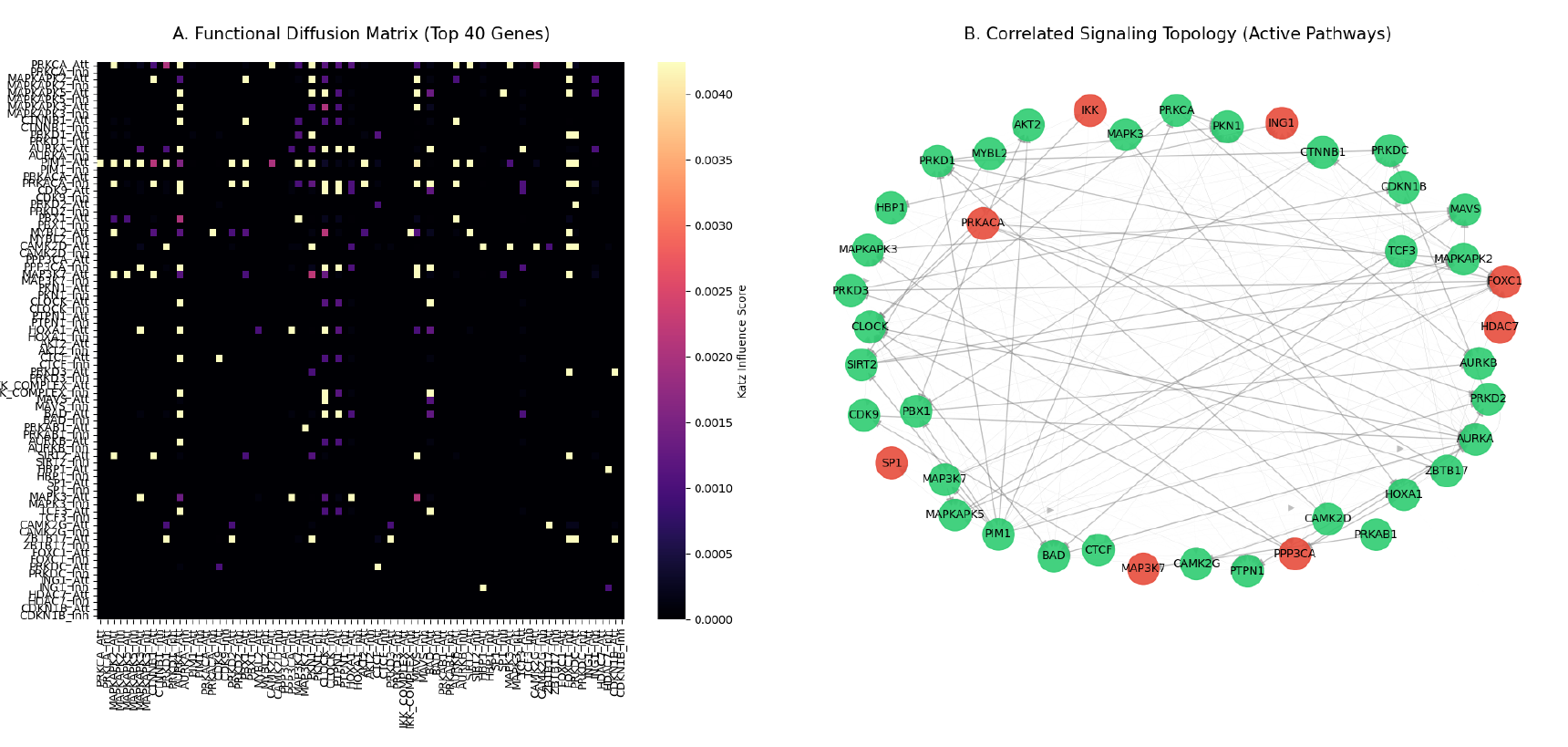}
\caption{\textbf{Graph Topology}.
(a) On the left it's shown a sample of genes and their activation/inhibition in path terms built using Katz techinque. The structure is used from Figure \ref{example_katz}. (b) On the right the graph built on the same nodes of the matrix on the left with colors defining if they are activated (green) or inhibited (red).
\label{fig:KatzDiffusion}}
 \end{center}
\vspace{-3mm}
\end{figure}

Comparing two adjacency matrix of a mechanistic graph only considers the presence of same edges within the graphs but not the entire topology comparison, like possible paths. To solve this problem and build a method that takes into account this structure the idea is to use the Katz diffusion matrix \cite{katz1953new}, defined as follows:

\begin{equation}
    M = (I - \delta A)^{-1} - I
\end{equation}

where $A$ is the unweighted adjacency matrix, $I$ is the identity matrix, and $\delta$ is the attenuation factor, constrained to be strictly less than the reciprocal of the largest eigenvalue of $A$ to ensure convergence. 

\begin{figure}[h]
    \centering
    \begin{tikzpicture}[
        node distance=2cm,
        gene/.style={circle, draw=black, thick, minimum size=1cm}
    ]
        \node[gene] (g1) {$g_1$};
        \node[gene] (g2) [right=of g1] {$g_2$};
        \node[gene] (g3) [right=of g2] {$g_3$};
        \node[gene] (g4) [right=of g3] {$g_4$};

        \draw[-{Stealth}, thick] (g1) -- (g2) node[midway, above] {$\delta$};
        \draw[-{Stealth}, thick] (g2) -- (g3) node[midway, above] {$\delta$};
        \draw[-{Stealth}, thick] (g3) -- (g4) node[midway, above] {$\delta$};

        \draw[-{Stealth}, bend left, dashed, gray] (g1) to node[midway, above] {$\delta^2$} (g3);
        \draw[-{Stealth}, bend left=40, dashed, lightgray] (g1) to node[midway, above] {$\delta^3$} (g4);
        
    \end{tikzpicture}
    \caption{A representation of a linear signal cascade. Solid arrows represent direct interactions (matrix A), whilst dashed arrows represent the indirect influence captured by the diffusion matrix M.}
    \label{fig:example_katz}
\end{figure}
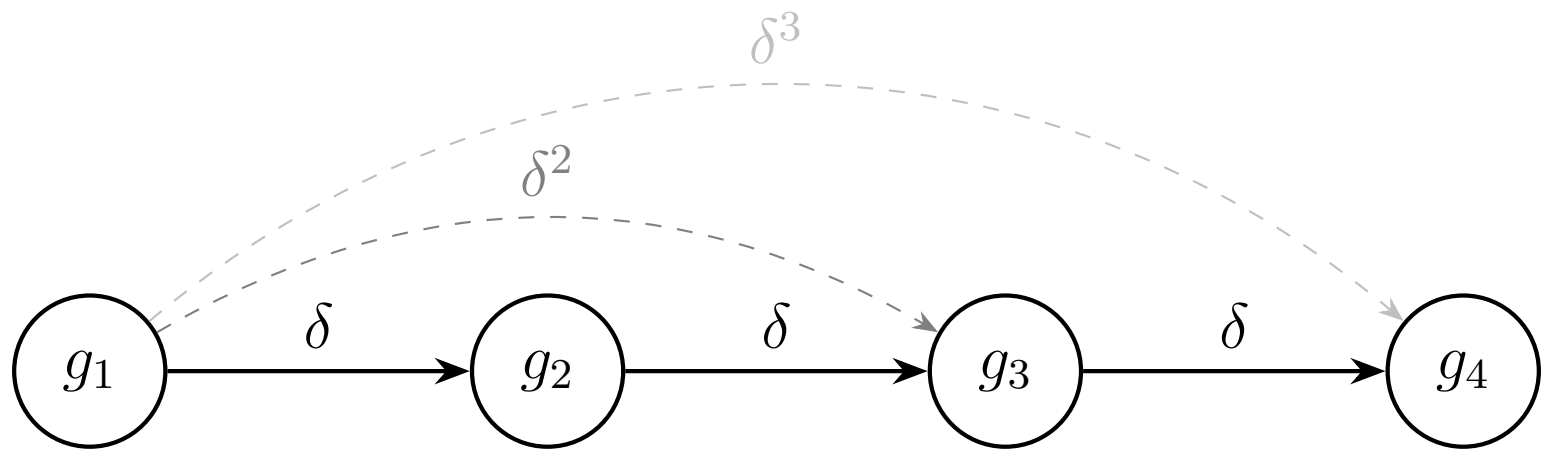
The diffusion matrix $M$ is computed purely on the structural network topology. This methodological choice ensures that the resulting matrix remains Positive Semi-Definite. Crucially, the biological logic is not lost due to the choice made before; rather, it is explicitly captured by the node alignment terms. Since the functional state of a node inherently reflects the state outcome of the signed regulatory interactions, evaluating structural capacity (Katz) and functional outcome (node states) independently avoids redundant penalization while preserving the full biological context \cite{chen2016katz}.

To compare the two matrices we decided to adopt the Frobenius inner product. Using this idea we're able to project the matrices in an Hilbert space, where the trace serves as a measure of overall alignment between signal structures. Unlike simple local metrics, the trace captures the consistency of the entire network architecture:

\begin{equation}
    S(\mathcal{G}_i, \mathcal{G}_2) = \langle M_1, M_2 \rangle_F = \sum_{i,j} (M_1)_{ij} (M_2)_{ij} = \text{Tr}(M_1^T M_2)
\end{equation}

Since the Katz matrix $M$ encodes the influence between genes on paths of infinite length, the trace of their product, $\text{Tr}(M_1^T M_2)$, acts as a weighted counter of shared functional pathways \cite{borgwardt2005protein}. In our $2N \times 2N$ representation, a pathway contributes to similarity only if it preserves both topology and activation state. This ensures that two patients are considered similar only if they activate the same molecular mechanisms. 

The proposed similarity measure $S(\mathcal{G}_1, \mathcal{G}_2)$ is formally defined as the Frobenius inner product between the state aware Katz diffusion matrices. This formulation is mathematically equivalent to a linear graph kernel, which satisfies the conditions of a Mercer’s Kernel \cite{hofree2013network}.

To build a measure that is included in a known interval let's normalize as follows:

\begin{equation}
    K(\mathcal{G}_1, \mathcal{G}_2) = \frac{\text{Tr}(M_1^T M_2)}{\sqrt{\text{Tr}(M_1^T M_1) \cdot \text{Tr}(M_2^T M_2)}}
\end{equation}

Since the Frobenius inner product is the canonical inner product in the space of matrices, the resulting similarity matrix $\mathbf{K}$ is by construction Symmetric Positive Semi-Definite. According to Mercer’s Theorem, any SPSD kernel implicitly defines a mapping $\phi: \mathcal{X} \to \mathcal{H}$ from the input space of graphs to a high dimensional Reproducing Kernel Hilbert Space (RKHS) \cite{hofmann2008kernel}, such that $K(\mathcal{G}_i, \mathcal{G}_j) = \langle \phi(\mathcal{G}_i), \phi(\mathcal{G}_j) \rangle_\mathcal{H}$.

While hierarchical clustering algorithms require a distance metric rather than a similarity measure, the existence of this inner product allows us to induce a valid Euclidean distance within the feature space $\mathcal{H}$. The distance between two signaling topologies is defined as the norm of their difference:

\begin{equation}
    d^2(\mathcal{G}_i, \mathcal{G}_j) = \|\phi(\mathcal{G}_i) - \phi(\mathcal{G}_j)\|^2_\mathcal{H}
\end{equation}

By expanding the dot product, we obtain the distance formula used for our clustering framework:

\begin{equation}
\label{equation:distance}
    d(\mathcal{G}_i, \mathcal{G}_j) = \sqrt{K(\mathcal{G}_i, \mathcal{G}_i) + K(\mathcal{G}_j, \mathcal{G}_j) - 2K(\mathcal{G}_i, \mathcal{G}_j)}
\end{equation}

This transformation ensures that the subsequent hierarchical Clustering operates on a mathematically sound metric space, where proximity reflects the degree of alignment between the functional signaling flows of the patients.

\section{Results}
\label{sec:RESULTS}

We analysed networks from 122 breast cancer patients, represented as mechanistic models inferred from transcriptomics, proteomics, and phosphoproteomics data and made available through the PatientProfiler tool \cite{lombardi2025patientprofiler}.

\begin{figure*}[h]
    \centering
    \begin{subfigure}[b]{0.5\textwidth}
        \centering
        \includegraphics[width=0.9\linewidth]{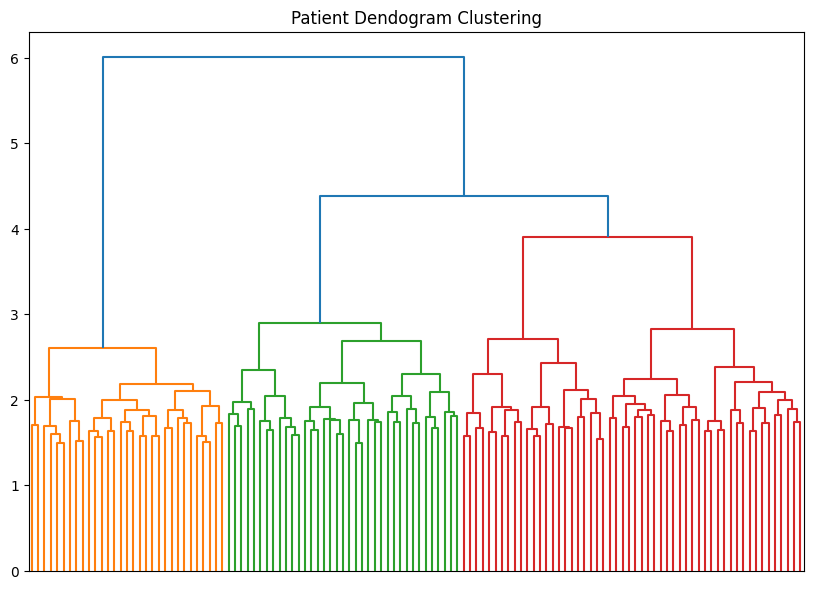}
        \caption{Hierarchical clustering Dendogram}
        \label{fig:hierarchical}
    \end{subfigure}%
    \hfill
    \begin{subfigure}[b]{0.5\textwidth}
        \centering
        \includegraphics[width=0.7\linewidth]{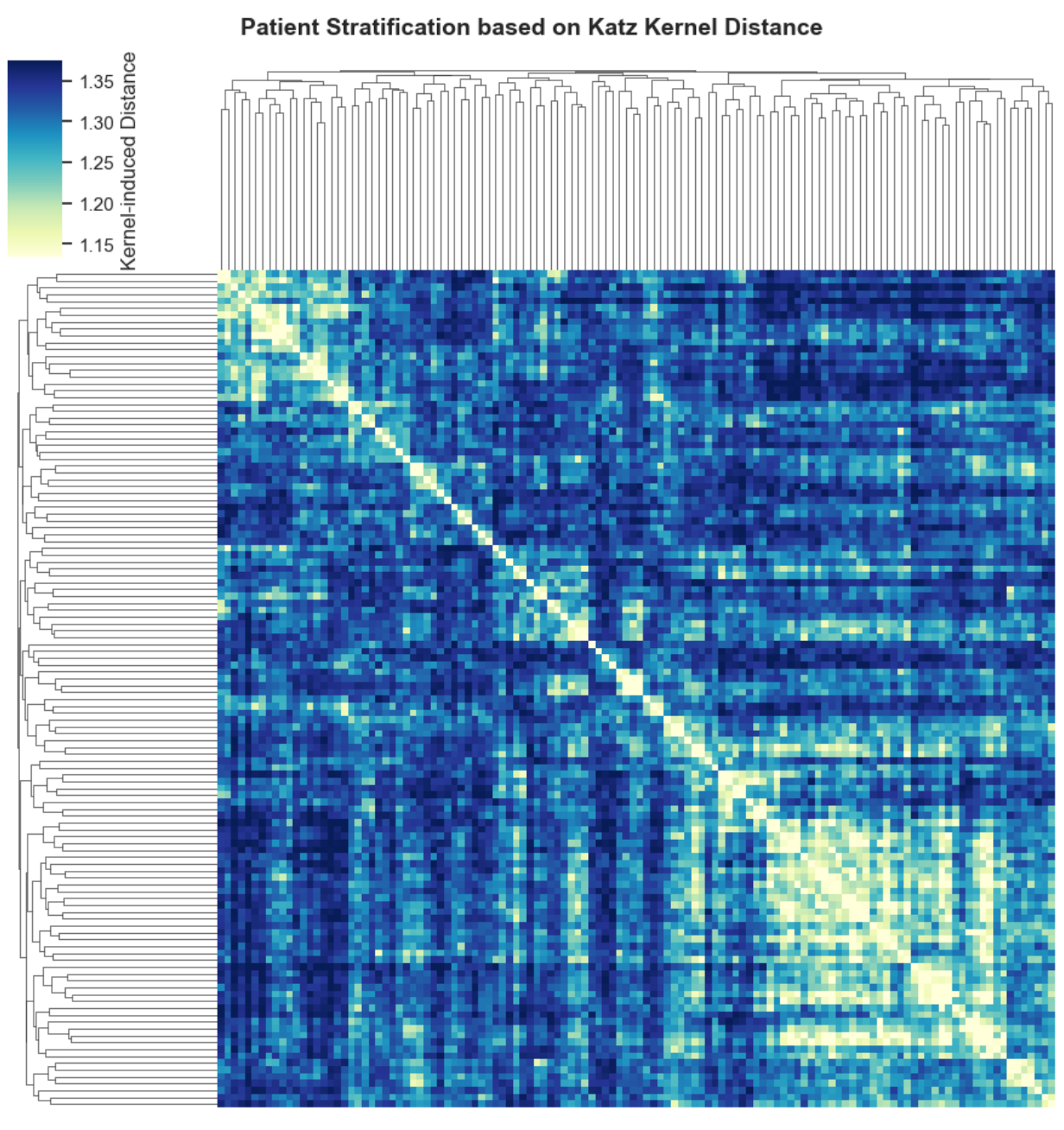}
        \caption{Katz Kernel Distance Matrix}
        \label{clustermap}
    \end{subfigure}%
    \vspace{-2mm}
    \label{fig:consensus_networks}
    \caption{Clusters Formation and Description}
\end{figure*}

The similarity measure described in Section \ref{sec:methods} is performed for every pair of patients. The results are then collapsed into a similarity matrix, from which we can compute a corresponding distance measure in Equation \ref{equation:distance}.  From the distance computed previously a hierarchical clustering is performed. From Figure \ref{fig:hierarchical}, the resulting dendrogram reveals a clear partition of the cohort into three stable subgroups, as evidenced by a  correlation and branch separation. This robustness is further confirmed by the clustered similarity heatmap, which displays modular blocks along the diagonal. The high internal consistency within each cluster, contrasted with the low inter cluster similarity, demonstrates that the Katz Kernel effectively captures some topological similarities in signaling architectures.

\begin{figure}[h]
    \centering
    \includegraphics[width=\linewidth]{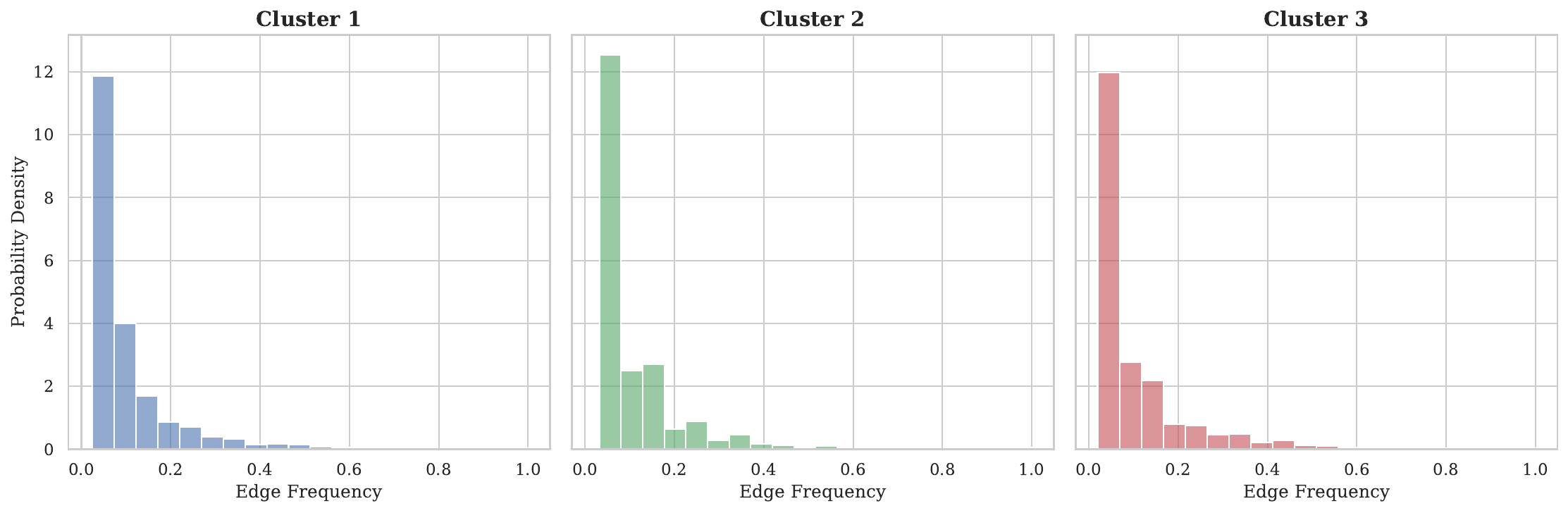}
    \caption{Statistical distribution of interaction frequencies across the identified patient clusters.}
    \label{fig:ppp}
\end{figure}

The statistical distribution of edge frequencies (Figure \ref{fig:ppp}) highlights a clear divergence in the internal consistency of the three groups. While Cluster 1 shows a broad distribution, indicating a heterogeneous signaling landscape with high intra patient variability, Clusters 2 and 3 are characterized by a sharp peak at high frequency values. This concentration suggests the presence of a highly conserved signaling backbone shared across the majority of patients in these subgroups. These results provide a quantitative basis for the clusters identified, showing that the Katz distance separates patients not only by similarity but also by the underlying rigidity of their signaling networks.

Once the robustness of the three clusters was visually confirmed by the block formed on the diagonal of the distance matrix (Figure \ref{fig:consensus_networks}), we proceeded to characterize their functional differences through consensus networks \cite{creixell2015}.

The functional identity of the three identified clusters is revealed in Figure \ref{fig:consensus_networks} through their consensus signaling networks, which were reconstructed by retaining interactions present in at least 50\% of the patients within each group. Cluster 1 exhibits a highly dense and interconnected regulatory core, suggesting a broad disregulation of signaling pathways where multiple hubs coordinate both activating and inhibitory signals. In contrast, Cluster 2 displays a more modular topology with a significant reduction in edge density, indicating a shift toward specific signaling axes rather than a global network overhaul. Finally, Cluster 3 represents a minimalist signaling profile where only the most robust and conserved interactions persist, highlighting a streamlined set of potential therapeutic targets. The marked differences in these topologies ranging from high complexity to sparse, specific modules provide clear evidence that the Katz kernel distance effectively discriminates between distinct biological states characterized by the rewiring of signaling architectures rather than mere fluctuations in gene expression levels.

\begin{figure*}[h]
    \centering
    \begin{subfigure}[b]{0.3\textwidth}
        \centering
        \includegraphics[width=\linewidth]{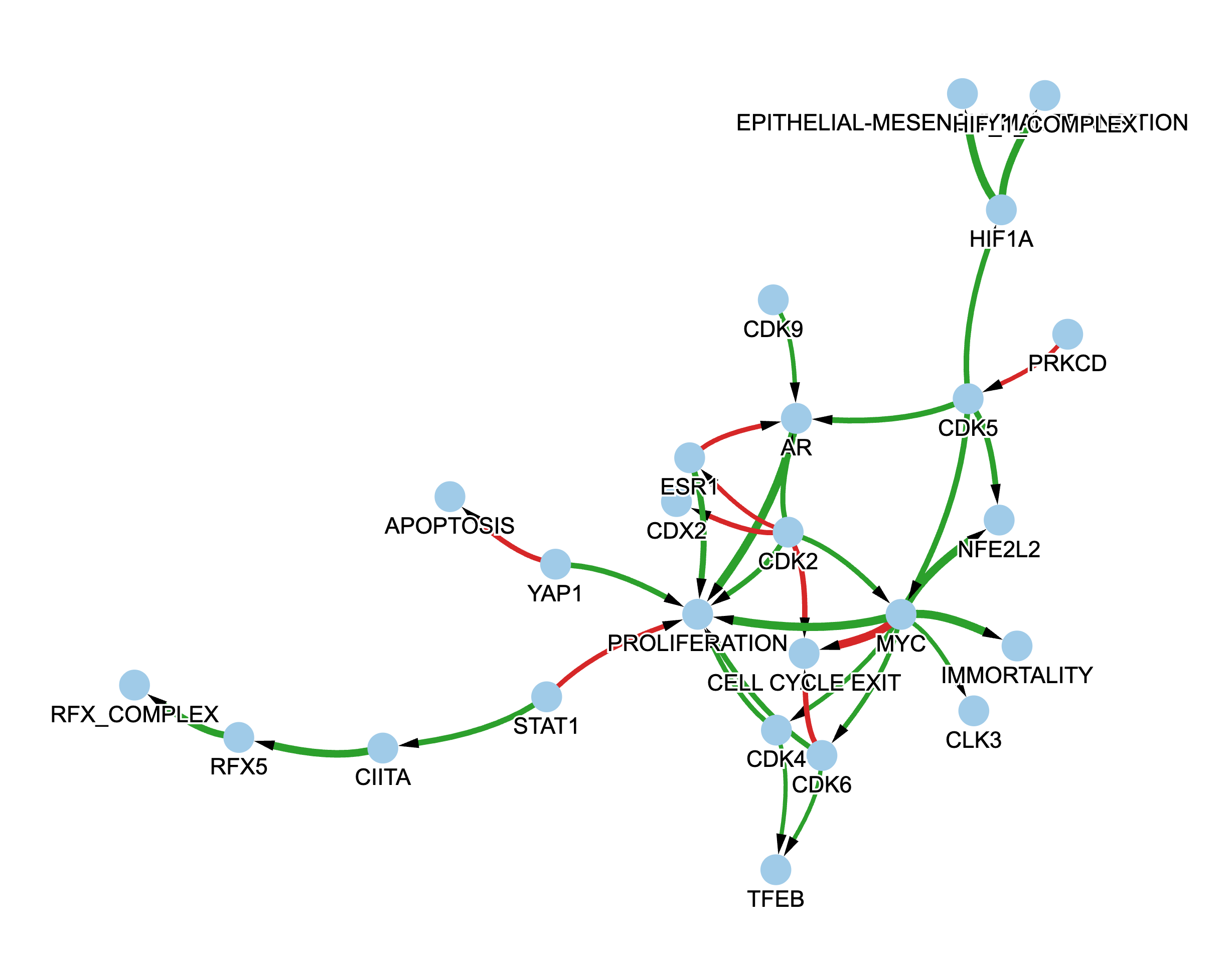}
    \end{subfigure}%
    \hfill
    \begin{subfigure}[b]{0.3\textwidth}
        \centering
        \includegraphics[width=\linewidth]{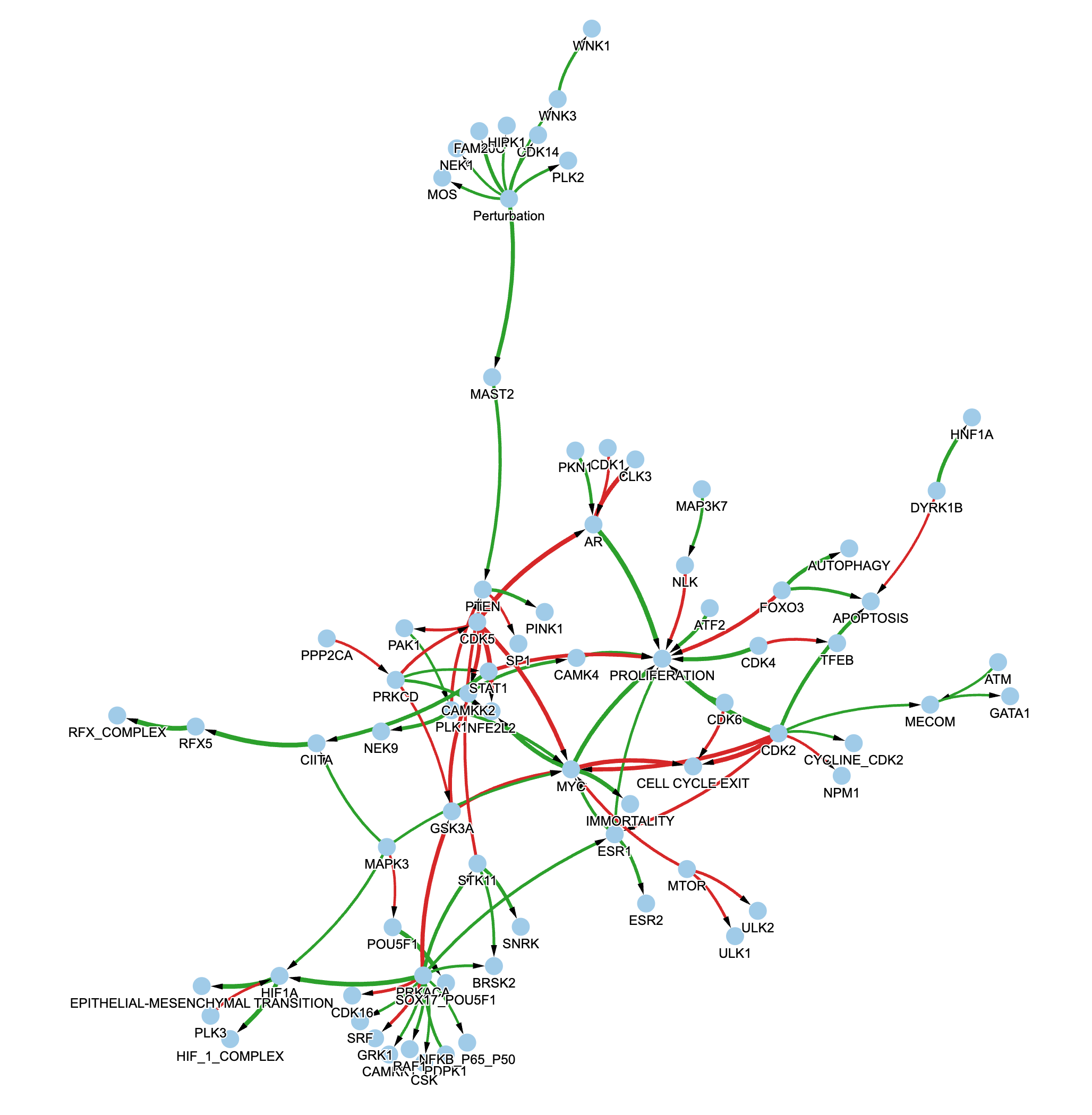}
    \end{subfigure}%
    \hfill
    \begin{subfigure}[b]{0.3\textwidth}
        \centering
        \includegraphics[width=\linewidth]{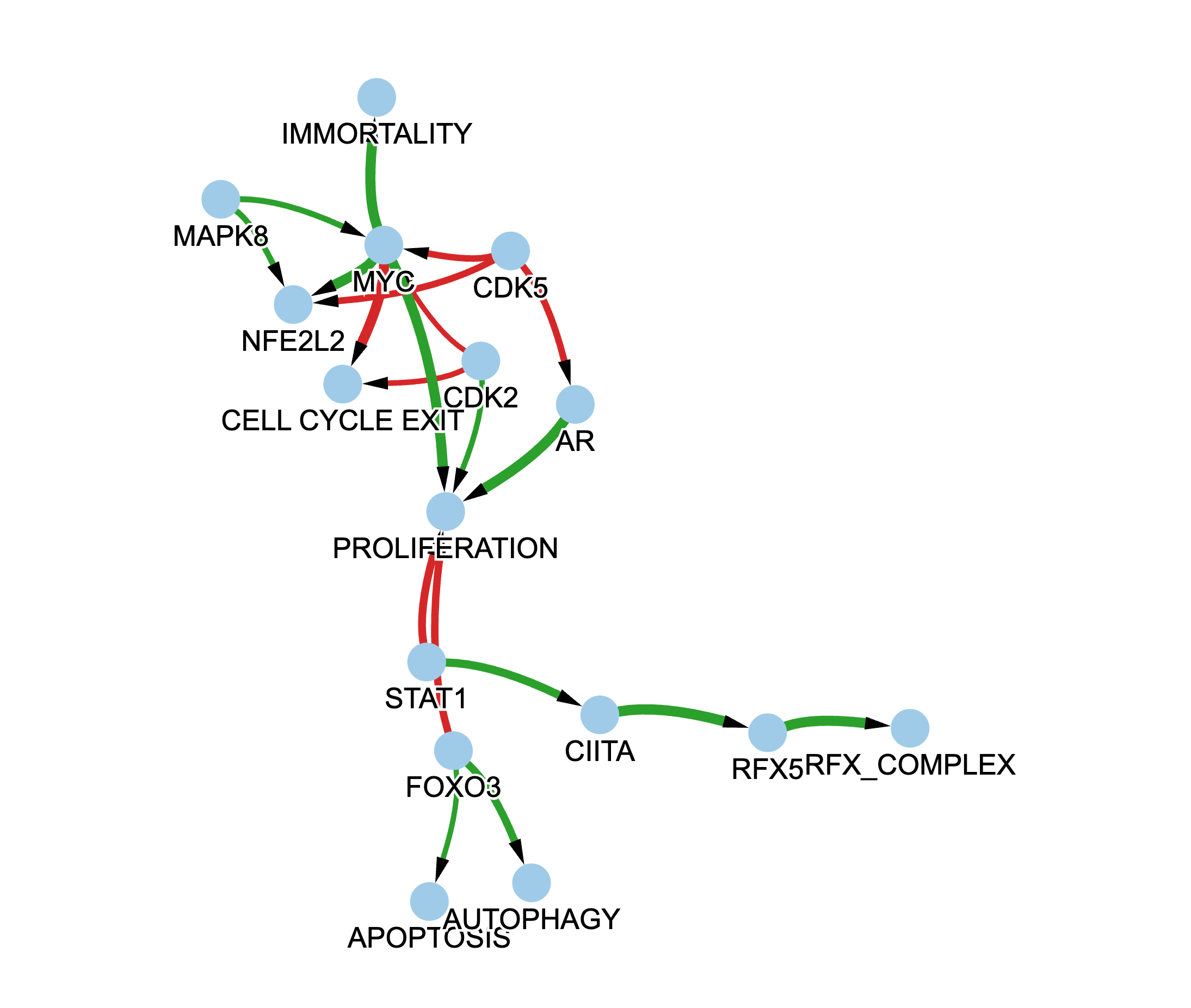}
    \end{subfigure}
    \vspace{-2mm}
    \caption{Consensus signaling networks for the three identified clusters (Cluster 1, 2, and 3 from left to right). Edges represent interactions present in at least 50\% of the patients. Green and red arrows denote activation and inhibition, respectively.}
    \label{fig:consensus_networks}
\end{figure*}

\section{Conclusion}
\label{sec:CONCLUSIONS}

In this work, we presented a novel framework for patient stratification based on signaling network topologies. By leveraging the Katz kernel to compute a robust distance metric between patient specific graphs, we identified three clusters characterized by different architectures. Unlike some traditional clustering methods, this approach allows to capture the complexity of signal transduction.

The analysis of consensus networks revealed a spectrum of non regulation : from the high density, interconnected noise of Cluster 1 to the streamlined, modular pathways of Cluster 3. The diagonal structure observed in the similarity heatmap provides strong empirical validation of the kernel’s ability to isolate biological states. These findings suggest that patients within the same clinical cohort could be driven by fundamentally different signaling logics, potentially explaining differential responses to targeted therapies. Future developments will focus on integrating these topological strata with clinical outcome data to assess the predictive power of stratification and test it with other kernels based on this kind of architecture.

\section*{Acknowledgments}
\label{sec:ACKNOWLEDGMENTS}
This work was supported by the Istituto Superiore di Sanità (ISS) through an external PhD fellowship granted by Sapienza University of Rome.

\footnotesize
\bibliographystyle{unsrt}
\bibliography{bibliography_CIBB_file.bib} 

@article{borgwardt2005protein,
  title={Protein function prediction via graph kernels},
  author={Borgwardt, Karsten M and Ong, Cheng Soon and Sch{\"o}nauer, Stefan and Vishwanathan, SVN and Smola, Alex J and Kriegel, Hans-Peter},
  journal={Bioinformatics},
  volume={21},
  number={suppl\_1},
  pages={i47--i56},
  year={2005},
  publisher={Oxford University Press}
}

@article{lombardi2025patientprofiler,
  title={PatientProfiler: A network-based approach to personalized medicine},
  author={Lombardi, Veronica and Di Rocco, Lorenzo and Meo, Eleonora and Venafra, Veronica and Di Nisio, Elena and Perticaroli, Valerio and Nicolaeasa, Mihail Lorentz and Cencioni, Chiara and Spallotta, Francesco and Negri, Rodolfo and others},
  journal={bioRxiv},
  pages={2025--01},
  year={2025},
  publisher={Cold Spring Harbor Laboratory}
}

@article{hofmann2008kernel,
  title={Kernel methods in machine learning},
  author={Hofmann, Thomas and Sch{\"o}lkopf, Bernhard and Smola, Alexander J},
  year={2008}
}

@article{hofree2013network,
  title={Network-based stratification of tumor mutations},
  author={Hofree, Matan and Shen, John P and Carter, Hannah and Gross, Andrew and Ideker, Trey},
  journal={Nature methods},
  volume={10},
  number={11},
  pages={1108--1115},
  year={2013},
  publisher={Nature Publishing Group US New York}
}

@article{katz1953new,
  title={A new status index derived from sociometric analysis},
  author={Katz, Leo},
  journal={Psychometrika},
  volume={18},
  number={1},
  pages={39--43},
  year={1953},
  publisher={Springer-Verlag}
}

@article{creixell2015,
  title={Kinome-wide decoding of network-rewiring mutations in cancer},
  author={Creixell, Pau and others},
  journal={Cell},
  volume={163},
  number={1},
  pages={202--217},
  year={2015}
}

@article{chen2016katz,
  title={Katz measures based prediction of disease-gene associations},
  author={Chen, Xing and others},
  journal={IEEE/ACM transactions on computational biology and bioinformatics},
  volume={14},
  number={4},
  pages={928--937},
  year={2016}
}
\normalsize

\end{document}